\documentclass[%
preprint,
superscriptaddress,
amsmath,amssymb,
aps,
]{revtex4-2}

\usepackage{graphicx}
\usepackage{dcolumn}
\usepackage{bm}

\usepackage{color}
\usepackage{amsmath}
\usepackage[caption=false]{subfig}

\begin{document}
\title{Non-equilibrium dynamics of bacterial colonies — growth, active fluctuations, segregation, adhesion, and invasion}
\author{Kai Zhou}
\affiliation{
Institute of Biological Information Processing
Theoretical Physics of Living Matter (IBI-5/IAS-2), Research center Juelich, Juelich, Germany}
\author{Marc Hennes}
\affiliation{Institute for Biological Physics and Center for Molecular Medicine Cologne, University of Cologne, Cologne, Germany}
\author{Berenike Maier}
\affiliation{Institute for Biological Physics and Center for Molecular Medicine Cologne, University of Cologne, Cologne, Germany}
\author{Gerhard Gompper}
\affiliation{
Institute of Biological Information Processing
Theoretical Physics of Living Matter (IBI-5/IAS-2), Research center Juelich, Juelich, Germany}
\author{Benedikt Sabass}
\email{B.Sabass@lmu.de}
\affiliation{Institute for Infectious Diseases and Zoonoses, Department of Veterinary Sciences, Ludwig-Maximilians-Universitaet Munich, Munich, Germany}
\affiliation{
Institute of Biological Information Processing
Theoretical Physics of Living Matter (IBI-5/IAS-2), Research center Juelich, Juelich, Germany}

\begin{abstract}
Colonies of bacteria endowed with a pili-based self-propulsion machinery are ideal models for investigating the structure and dynamics of active many-particle systems. We study \textit{Neisseria gonorrhoeae} colonies with a molecular-dynamics-based approach. A generic, adaptable simulation method for particle systems with fluctuating bond-like interactions is devised. The simulations are employed to investigate growth of bacterial colonies and the dependence of the colony structure on cell-cell interactions. In colonies, pilus retraction enhances local ordering. For colonies consisting of different types of cells, the simulations show a segregation depending on the pili-mediated interactions among different cells. These results agree with experimental observations. Next, we quantify the power-spectral density of colony-shape fluctuations \textit{in silico}. Simulations predict a strong violation of the equilibrium fluctuation-response relation. Furthermore, we show that active force generation enables colonies to spread on surfaces and to invade narrow channels. The methodology can serve as a foundation for future studies of active many-particle systems at boundaries with complex shape.
\end{abstract}

\maketitle
\section{Introduction}
Bacterial colonies consisting of cells with nearly identical geometry and mechancial properties are uniquely suited for studying the non-equilibrium statistical mechanics of living matter~\cite{bonazzi2018intermittent, welker2018molecular, shaebani2020computational}. A well-established biological model system is the coccoid/diplococcoid bacterium \textit{Neisseria gonorrhoeae}. With a spherical cell body with a diameter of roughly $1\,\mu\mathrm{m}$, the bacterium forms colonies that 
are reminiscent of nonliving colloidal assemblies. However, bacteria grow and reproduce. Moreover, while colloidal assemblies are held together by passive attractive interactions, such as depletion forces, \textit{N.~gonorrhoeae} colonies are held together by extracellular filaments called type IV pili (T4P) that are cyclically elongated and retracted~\cite{craig2019type}. T4P are helical polymers consisting mainly of the major subunit PilE. Anchored in a transmembrane complex, T4P are isotropically displayed on the whole cell surface of \textit{N.~gonorrhoeae}~\cite{marathe2014bacterial}. Their elongation and retraction is driven by the dedicated ATPases PilF and PilT, respectively. PilF is required for pilus polymerization and PilT drives pilus retraction and depolymerization. During retraction, T4P are capable of generating high forces exceeding $100\,$pN~\cite{merz2000pilus, maier2002single}, which is 20 times higher than the force generated by muscle myosin and makes the T4P one of the strongest molecular machines known so far~\cite{zollner2019motor}, and the retraction proceeds with velocities up to $2\,\rm{\mu m}/ \rm{s}$~\cite{marathe2014bacterial, welker2018molecular}. When individual cells come into proximity of abiotic surfaces such as glass, cells can attach via T4P. Since \textit{N.~gonorrhoeae} generates multiple pili simultaneously, a tug-of-war between pili on different sides of the cell body ensues. On glass surfaces, the average detachment force is an order of magnitude smaller than the maximum force generated by pili. Therefore, the tug-of-war leads to a random walk of individual bacteria on surfaces~\cite{skerker2001direct,holz2010multiple, marathe2014bacterial,zaburdaev2014uncovering, ponisch2017multiscale, ponisch2019bacterial}. In aerobic environments, individual cells form colonies. In colonies, cells are held together by T4P, which produce time-dependent attractive interactions among bacteria~\cite{kurre2012oxygen,dewenter2015oxygen}. The fact that this interaction is caused by time-dependent non-equilibrium forces affects the shape, dynamics, and sorting behavior of bacterial colonies. \textit{N.~gonorrhoeae} mutants without T4P cannot aggregate into colonies~\cite{taktikos2015pili}. The strength of cell-cell attraction is affected by T4P post-translational modifications and can be controlled by inhibiting or activating different steps of the pilin glycosylation pathway~\cite{zollner2019type}. 

The material properties of \textit{Neisseria} colonies have been characterized as liquid-like~\cite{bonazzi2018intermittent} with effective viscosities of $\eta \sim 350\,\mathrm{Pa\,s}$ for \textit{N.~gonorrhoeae}~\cite{welker2018molecular}. Microcolonies display properties that are partially reminiscent of droplets exhibiting an effective surface tension. Evidence for an effective surface tension is firstly the spherical shape of microcolonies formed by \textit{N.~gonorrhoeae} with retractile T4P~\cite{higashi2007dynamics}. Secondly, upon contact, two microcolonies fuse to form a sphere with larger radius~\cite{dewenter2015oxygen,taktikos2015pili, welker2018molecular}. Depending on the strength and activity of T4P interactions, initial fusion is however followed by slow coalescence of the two microcolonies that can take hours~\cite{ponisch2017multiscale, ponisch2018pili} and the mechanical response of colonies certainly contains elastic components on some time scales.

While material properties of \textit{N.~gonorrhoeae} colonies have been characterized~\cite{bonazzi2018intermittent}, some non-equilibrium effects resulting from active bacterial force-generation remain to be explored. In thermodynamic equilibrium, the velocity correlation measured in particle systems is generally proportional to the linear response with respect to a small perturbation, which is called a fluctuation-response relation~\cite{seifert2012stochastic}. The extent of the violation of this fluctuation-response relation in non-equilibrium states can be related to the rate of energy dissipation~\cite{harada2005equality,prost2009generalized}. For bacterial colonies, active force-generation by T4P changes the fluctuations of the conservative forces experienced by the cells and entails energy dissipation. Therefore, the fluctuation-response relation is expected to be violated in bacterial colonies under the premise that colonies can be described as physical particle system~\cite{gnesotto2018broken, mizuno2007nonequilibrium, netz2018fluctuation}. However, the frequency-characteristics and measurability of this violation are unknown. Thus, an aim of this work is to establish theoretical predictions regarding the non-equilibrium fluctuations of \textit{N.~gonorrhoeae} colonies.

Internally driven many-particle systems can also exhibit non-equilibrium phase transitions, which has been investigated for some classes of model systems~\cite{cates2015motility, speck2014effective,zottl2016emergent,digregorio2018full}. Notably, for active particles that undergo rotational motion and thus on average obey rotational symmetry, density fluctuations are Gaussian and the non-equilibrium phase transitions can be understood in a framework similar to equilibrium phase transitions~\cite{del2018energy,han2017effective}. For \textit{N.~gonorrhoeae} colonies, the average force generation by individual bacteria is presumably almost spherosymmetric and therefore it may be challenging to distinguish genuine non-equilibrium colony dynamics from dynamics that are also be observable in passive systems. However, it has been shown experimentally that 
mechanical forces govern the sorting of different cells during the early formation of \textit{N.~gonorrhoeae} colonies~\cite{oldewurtel2015differential}. Mutants with different T4P density and rupture forces of T4P-mediated adhesion spatially segregate inside colonies, suggesting a sorting process driven by pilus retraction that also depends on differential adhesiveness~\cite{oldewurtel2015differential, ponisch2018pili}. Self-sorting of \textit{Neisseriae} colonies has been studied experimentally by changing the post-translational modification of T4P, their activity, and computer simulations have been conducted~\cite{oldewurtel2015differential, bonazzi2018intermittent, ponisch2018pili}. 

In general, physical properties of large bacterial colonies are ideally studied with a combination of experiments, theory, and detailed computer simulations. Previous work includes simulations of the dynamics of single cells due to individual pili~\cite{marathe2014bacterial, zaburdaev2014uncovering,  ponisch2017multiscale, ponisch2019bacterial} and coarse-grained approaches or continuum theories for the description of \textit{Neisseria} colonies~\cite{kuan2021continuum}. Furthermore, multiscale simulations combining overdamped cell dynamics with stochastic pilus activity have shown great promise for the investigation of the behavior of \textit{Neisseria} colonies on different length scales~\cite{ponisch2017multiscale,bonazzi2018intermittent}. Mechanical forces in bacterial colonies are not only actively generated by T4P but also by cell growth and division. For the case of mammalian cells, tissue growth has been studied extensively with particle-based simulations where individual cells are represented as spheres~\cite{ranft2010fluidization,podewitz2015tissue,ganai2019mechanics}. The sphericity and growth dynamics assumed for cells with these models are also appropriate for simulating coccoid bacteria.

The present work is based on a code for multiscale simulation of colonies consisting of coccoid bacteria that employ T4P to generate active forces, while also growing and dividing. We employ here the highly parallel, classical molecular dynamics simulator LAMMPS~\cite{plimpton1995fast} and add a dedicated extension for the simulation of growing cells that interact with each other through elastic, retractable bonds representing pili. Using this model, we simulate the growth dynamics of colonies and the structural order of cells inside colonies. With appropriate parameterization, the simulation results are qualitatively consistent with experimental findings. Simulations results for cell segregation in colonies consisting of different mutants of the T4P machinery also agree qualitatively with experimental results. Furthermore, we predict a strong violation of the equilibrium fluctuation-response theorem for the colony shape and show that a colony invasion of narrow channels is driven by active pilus-mediated forces.


\section{Results and Discussion}
\subsection{Simulation of colonies of coccoid cells} 
\label{methods}
In our simulations, colonies are grown from individual cells through cell division. An individual cell is also called a coccus. A pair of dividing \textit{N.~gonorrhoeae} cells is called a diplococcus and has the shape of two partially overlapping spheres. During the growth, each diplococcus divides approximately with rate $\alpha$ into a pair of individual cocci, which in turn again become diplococci after some time. The fraction of dead cells in \textit{N.~gonorrhoeae} colonies is reported to be below $5\,\%$~\cite{welker2021spatio} and cell death is therefore neglected in our model. Each individual coccus is endowed with a repulsive potential modeling volume exclusion. In addition, cells experience dissipative forces resulting from relative motion of neighboring cells and thermal fluctuations. Each cell has a fixed number of pili. By modeling the pili as dynamic springs that can extend, retract, bind and unbind either with other pili or with the environment, we faithfully represent the stochatic nature of cell-generated forces.

\subsubsection{Cell geometry and bacterial growth}
All simulations are conducted in a three-dimensional, Cartesian coordinate system. Individual cells are modeled as soft spheres with radius $R$. The position of the center of bacterium $i$ is denoted by $\mathbf{r}_{i}$. The vector between a pair of bacteria with indices $(i,j)$ is denoted by $\mathbf{r}_{ij}=\mathbf{r}_i-\mathbf{r}_j$ and their distance is $r_{ij}=\left| \mathbf{r}_{ij} \right|$.

\begin{figure}[htp]
\includegraphics[width=0.7\columnwidth]{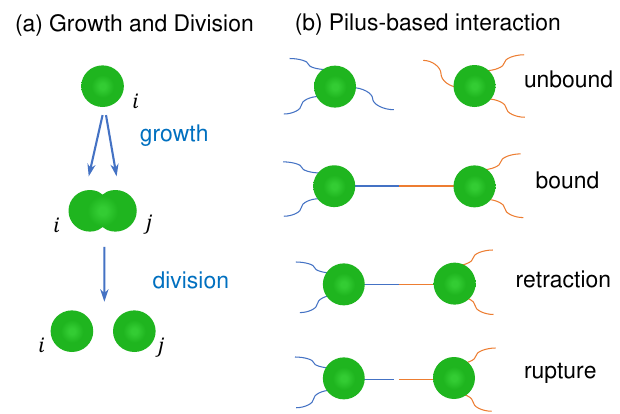} 
	\caption{Schematic representation of the two active processes occurring in the simulations: (a)~A bacterium grows into a diplococcus and then divides into two individual bacteria. (b)~Two cells bind to each other via pili. Subsequently, pili are retracted by the bacteria leading to a force build-up. The bond connecting the pili ruptures stochastically in a force-dependent manner. 
		}
		\label{simsketch}
\end{figure}

The division of a cell with index $i$ is modeled by insertion of a second sphere with index $j$ on top of $i$ so that the excluded volume remains the same during insertion, see Fig.~\ref{simsketch}(a). The pair of cells is initially connected by an elastic spring with time-dependent rest length $l(t)$, thus forming a diplococcus. The initial orientation of the vector connecting the cell pair is chosen randomly. Then, growth of the diplococcus is simulated by increasing the rest length of the spring that connects the cell pair as
\begin{equation}
\frac{{\rm d} l(t)}{{\rm d}t} = \frac{\alpha}{\nu_{\mathrm{r}}},
\label{eq_dldt}
\end{equation}
with a rate constant $\alpha$ and a constant with units of length $1/\nu_{\mathrm{r}}$. This linear growth of the long axis of the diplococcus with time is consistent with experimental data~\cite{westling1977growth}. Note that the geometry of the growth of the diplococcus implies that the volume growth rate increases with size of the individual diploccus. As a generalization, one could also include a growth rate $\alpha$ that explicitly depends on the size of the diplococcus to represent an arbitrary volume-dependency of the growth on the single-cell level~\cite{mir2011optical}. A further complication that has not been included in the model, for simplicity, is that the direction of growth and division in \textit{Neisseria} likely follows a complex pattern determined by alternating perpendicular division planes~\cite{westling1977growth}. Once the length $l(t)$ reaches a threshold $l_{\rm t}$, the connecting spring is removed and the two spheres are treated as individual cocci. Instantaneous forces acting on either of the two cocci during their separation are equally distributed among the two cocci to ensure momentum conservation. The time between creation of a diplococcus and separation of the two daughter cells is given by $t_{\mathrm{r}} = l_{\mathrm{t}} \nu_{\mathrm{r}}/\alpha$.

After separation of a diplococcus, the two individual cocci do not become diplococci instantaneously. Rather, individual cells are turned into diplococci at a constant rate per cell, for which we employ for simplicity the same rate constant $\alpha$ that also appears in Eq.~(\ref{eq_dldt}). This means that the separation of a diplococcus (division) is followed by a random refractory time, which prevents an unphysical synchronization of the division events in the simulations. 

The growth and division model employed in this work is primarily motivated by its simplicity and numerical stability. For several bacterial organisms, experimental studies have demonstrated that homeostasis of cell size can be explained by a phenomenological ``adder rule'', whereby cells increase by a constant volume each generation, regardless of initial size at division~\cite{campos2014constant,amir2014cell,willis2017sizing}. Thereby, the volume increment during each generation sets the division time and division times obey a Gaussian distribution. For our model, however, we choose to keep the sizes of the spheres always the same to enable a robust parametrization. This choice produces a natural scale for cell division and insofar determines the division times. The additive noise in bacterial division times that reportedly results in their Gaussian distribution~\cite{amir2014cell} is represented in our model approximately by the random refractory time between pair separation and creation of daughter cells in the next generation.

\subsubsection{Dynamics of type IV pili}
Each cell is assumed to have a constant number of pili, typically around $7$, see Fig.~\ref{simsketch}(b) and Tab.~\ref{table02}. A pilus of bacterium $i$ is assumed to bind with a rate $k_{\rm {bind}}$ to one pilus of a neighboring bacterium $j$. For binding, the distance between $i$ and $j$, $r_{ij}$, is required to be less than a cutoff distance $d_{\rm {bind}}$. This cutoff distance ensures that only bacteria bind to each other when they are in proximity to each other. In some simulations, a distance-based criterion and a Voronoi tessellation are combined to limit pilus interactions only to immediate neighbors that have a distance from each other that is smaller than the cutoff for pilus binding. For simplicity, we assume that pilus-based forces act along the straight lines connecting the centers of cell pairs. It is assumed that two bacteria can only have one pair of pili adhering to each other. Likewise, bundling of pili~\cite{marathe2014bacterial} is also neglected due to its unknown role in cell colonies. Pili of the two cells in a diplococcus do not bind to each other. The pilus-based cell-cell connection is modeled as a spring connecting the centers of two cells. The rest length of the pilus connecting two cells with indices $i$ and $j$ is denoted by
$L_{ij}$. The force exerted on the pair of cells is purely attractive and given by
\begin{equation}
f_{ij}^p =  {\rm{min}}\left[0, -k[r_{ij}(t)-L_{ij}(t)]\right],
\end{equation}
where $k$ is the pilus' spring constant. Once the pilus is bound, it is assumed to retract. Thus, the effective pilus dynamics employed for our model exclude non-retracting pili that form passive bonds among cells, see also earlier work~\cite{zaburdaev2014uncovering}. Pilus retraction leads to a continuous shortening of its rest length as
\begin{equation}
L_{ij}(t) = {\rm{max}}\left[2R, r_{ij}(0)- \int_{0}^{t}{v_{\rm {re}}(t)}{\rm d}t\right],
\end{equation}
where $v_{\rm {re}}$ is the force-dependent retraction velocity of pili. To describe the force-velocity relationship for T4P retraction motors~\cite{maier2002single}, we employ the linearized relation
\begin{equation}
v_{\rm {re}}(t) = {\rm{max}}\left[0, v_{\rm {re}}(0)\left(1-\frac{f_{ij}^p}{f_{\rm s}}\right)\right],
\end{equation}
where the stall force $f_{\rm s}$ represents the maximal force a retracting pilus can generate. Furthermore, it is assumed that the bonds between the pili rupture under stress with a force-dependent rate as
\begin{equation}
\gamma_{\rm {rupt}}=\frac{1}{t_1 e^{-f_{ij}^p/F_{\rm {c,1}}} + t_2 e^{-f_{ij}^p/F_{\rm {c,2}}}}, 
\end{equation}
where $t_1$ and $t_2$ are two characteristic rupture time, $F_{\rm {c,1}}$ and $F_{\rm {c,2}}$ are two characteristic rupture force.
To simplify the analysis of the colony dynamics, we employ an idealized pilus rupture rate starting from the section ``Active phase segregation in mixed colonies'' in all following sections of the Results and Discussion. The idealized pilus rupture rate is 
\begin{equation}
\gamma_{\rm {rupt}}=k_{\rm {rupt}} e^{f_{ij}^p/F_{\rm {rupt}}}, 
\end{equation}
where $k_{\rm {rupt}}$ is the pilus rupture rate without loading and $F_{\rm {rupt}}$ is a characteristic rupture force. Related models of pilus dynamics have been employed previously~\cite{marathe2014bacterial,zaburdaev2014uncovering,ponisch2017multiscale, simsek2019substrate}.

\subsubsection{Dynamics of bacteria}
For simulating the cell dynamics, we employ an algorithm similar to dissipative particle dynamics (DPD)~\cite{espa:95pr, groot1997dissipative}, where we assume a soft repulsion between cells, a frictional force proportional to the relative velocity of neighboring cells, and thermal noise forces that satisfy the Einstein relation. Underdamped equations of motion for every cell $i$ with mass $m_i$, position $\mathbf r_i$, velocity $\mathbf v_i$, and force $\mathbf f_i$ are assumed as
\begin{align}
    \frac{{\rm d} \mathbf{r}_i}{{\rm d}t}=\mathbf{v}_i, & &   m_i \frac{{\rm d} \mathbf{v}_i}{{\rm d}t}=\mathbf{f}_i.
\end{align}
The force acting on each pair of cells consists of conservative forces $\mathbf{F}_{ij}^c$, dissipative forces $\mathbf{F}_{ij}^d$, thermal fluctuations $\mathbf{F}_{ij}^r$ and forces from active pilus retraction $\mathbf{F}_{ij}^p$. Overall, the sum of these forces is 
\begin{equation}
\mathbf{f}_i = \sum_{j\neq i}(\mathbf{F}_{ij}^c+\mathbf{F}_{ij}^d+\mathbf{F}_{ij}^r+\mathbf{F}_{ij}^p).
\end{equation}
For defining the individual force terms, we employ the vector between the centers of masses $\mathbf{r}_{ij} = \mathbf{r}_{i}-\mathbf{r}_{j}$ and the unit vector pointing towards cell $i$ denoted by $\widehat{\mathbf{r}}_{ij}=\mathbf{r}_{ij}/r_{ij}$. 

The conservative force acting between pairs of unbound bacteria is 
\begin{equation}
\mathbf{F}_{ij}^c = \begin{cases}
a_{0}(1-r_{ij}/d_{\rm {con}})  \widehat{\mathbf{r}}_{ij} & (r_{ij}< d_{\rm {con}})\\
0 & (r_{ij}\geq d_{\rm {con}})
 \end{cases} ,
\end{equation}
where $a_{0}$ is the maximum conservative force between bacterium $i$ and $j$, the cutoff distance for the repulsive cell-cell interaction is denoted by $d_{\rm {con}}=2 R$. For a diploccus consisting of two spheres, the conservative force due to growth is 
\begin{equation}
\mathbf{F}_{ij}^c = a_{\rm {growth}} (l_i-r_{ij}) \widehat{\mathbf{r}}_{ij} ,
\end{equation}
where $a_{\rm {growth}}$ is the elastic constant of the spring connecting the two cells of a diplococcus.
The dissipative and random forces are, respectively, given by
\begin{eqnarray}
\mathbf{F}_{ij}^d&=&-\gamma \omega^D(r_{ij})(\widehat{\mathbf{r}}_{ij}\cdot\mathbf{v}_{ij})\widehat{\mathbf{r}}_{ij}, \\
\mathbf{F}_{ij}^r&=&\sqrt{2 \gamma  k_{\rm B}T } \omega^R (r_{ij}) \theta_{ij} \widehat{\mathbf{r}}_{ij},
\end{eqnarray}
where $\gamma$ is a friction coefficient, $\omega^D$ and $\omega^R$ are distance-dependent weight functions, $ k_{\rm B}$ is the Boltzmann constant, $T$ is the ambient temperature and $ \theta_{ij}=\theta_{ji}$ is a random number drawn from a Gaussian distribution with zero mean and unit variance. For the distance-dependence of the friction force we choose
\begin{equation}
\omega^D(r)=[\omega^R(r)]^2 = \begin{cases}
(1-r_{ij}/d_{\rm {dpd}})^2 & (r_{ij}< d_{\rm {dpd}})\\
0 & (r_{ij}\geq d_{\rm {dpd}})
\end{cases} ,
\end{equation}
where $d_{\rm {dpd}}$ is the cutoff distance for dissipative and random forces. Finally, the forces resulting from retraction of pili are given in their vectorial form by
\begin{equation}
\mathbf{F}_{ij}^p = f_{ij}^p \widehat{\mathbf{r}}_{ij}.
\end{equation}
Note that we do not consider the torques generated by T4P between pairs of cells, which, however has been incorporated in related models of others~\cite{ponisch2017multiscale}.

\subsection{Exponential colony growth}
\label{formation}
\begin{figure}[!htp]
	\includegraphics[width=1.0\columnwidth]{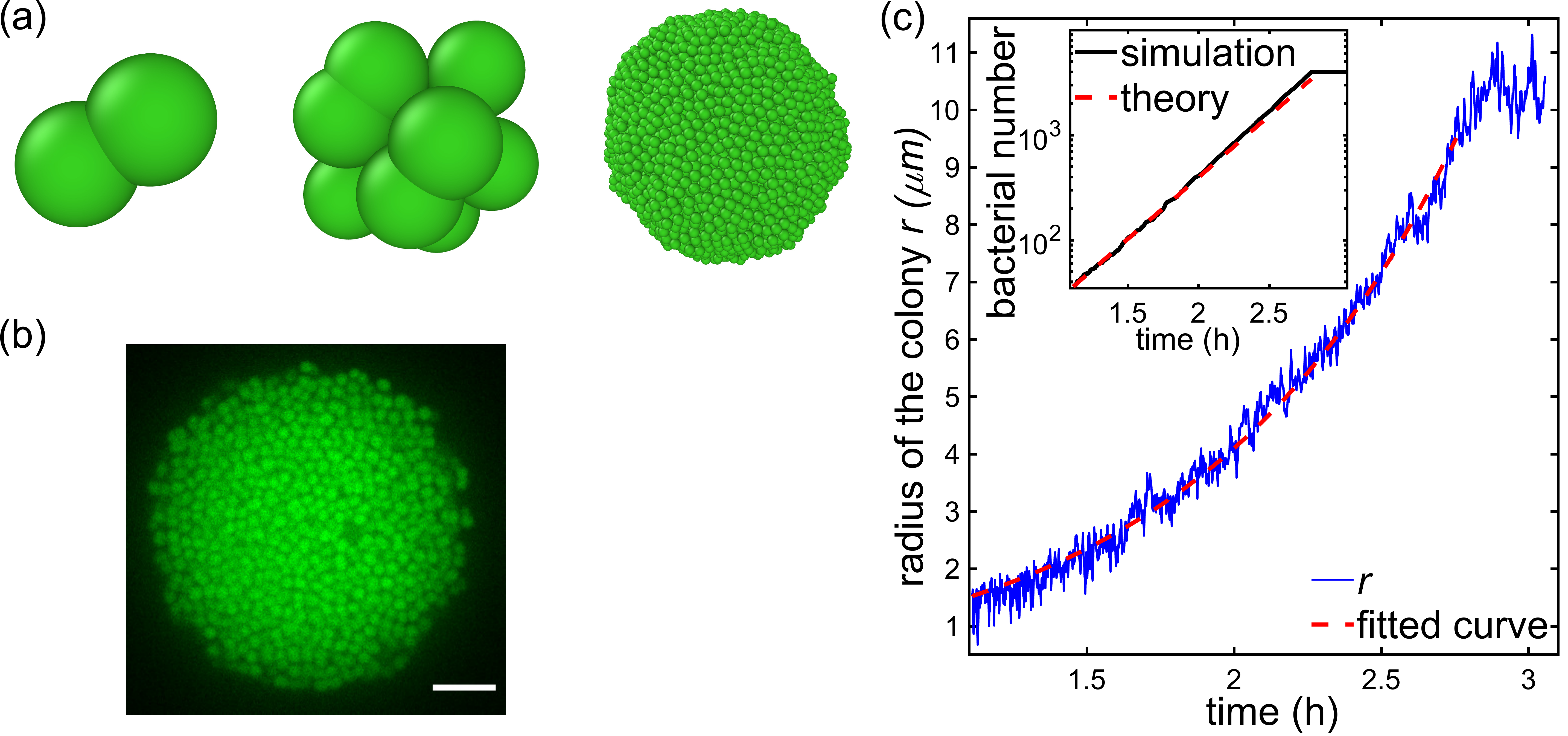}
	\caption{Simulations of growing colonies. (a)~Simulation snapshots showing colony formation.
		(b)~Confocal section of a microcolony of fluorescently labeled \textit{N.~gonorrhoeae}. Scale bar: $5\,\mu$m.
		(c)~The colony radii increase exponentially with time during the simulations. Growth and division are switched off before the end of the simulation. Inset: the number of cells increases exponentially with time and simulation results agree with an analytical expression.}
	\label{bacterianumber} 
\end{figure}

Figures~\ref{bacterianumber} shows simulation results for colony growth. As for experimental systems, colonies approximately maintain a spherical shape during growth. In the simulations, both the number of bacteria and the colony radius increase exponentially with time. Colony radii are quantified by measuring the distance between the center of mass and cells on the boundary at a fixed polar angle and azimuth. Since we have not taken into account a position-dependence of nutrient availability inside colonies, we expect to see such growth dynamics in experiments only for small cell colonies in rich media. Experimentally, an exponential increase of the radii of \textit{N.~gonorrhoeae} colonies has been observed for about three hours during the initial growth of young colonies~\cite{welker2021spatio}. 

Next, analytical formulas are derived for the simulated growth dynamics. The cell-growth simulations are based on the assumption of two growth phases - consisting of single cocci and diplococci. The advantage of this two-phase model is that it allows the introduction of a controllable randomization of division events and thus the avoidance of artificial division synchronization. A single coccus can divide to form a diplococcus, which is a random event that occurs with rate $\alpha$. The resulting diplococcus cannot divide immediately but grows on average for a time $t_{\rm r}$ until it separates into two single cocci that can then divide. We denote the average number of all cells forming the cocci and diplococci by $N(t)$. The average number of cells that are single cocci is denoted by $N_{\mathrm{c}}(t)$. Since only the single cocci are assumed to divide, the overall number of bacteria is determined by
\begin{equation}
\frac{{\rm d} N(t)}{{\rm d}t} = N_{\mathrm{c}}(t)\alpha.
\label{eq:Nc_division}
\end{equation}
We next consider the governing equation for the number of single cocci $N_{\mathrm{c}}(t)$, which increases at time $t$ through separation of diplococci. The separating diplococci, in turn, were formed at time $t-t_{\mathrm{r}}$ through division of single cocci. Hence, the increase of single cocci at time $t$ is given by $2 \alpha N_{\mathrm{c}}(t-t_{\mathrm{r}})$, where the factor $2$ results from cell doubling during division. Simultaneously, the number of single cocci is reduced through formation of diplococci with rate $ \alpha N_{\mathrm{c}}(t)$. Overall, we obtain
\begin{equation}
\frac{{\rm d} N_{\mathrm{c}}(t)}{{\rm d}t} = 2 \alpha N_{\mathrm{c}}(t-t_{\mathrm{r}})- \alpha N_{\mathrm{c}}(t).
\label{eq:Nc_fraction}
\end{equation}
Growth is assumed to obey an exponential time dependence and the ansatz $N_{\mathrm{c}}(t) = N_{\mathrm{c}}(0) e^{\alpha p t}$ with a constant $p$ is inserted into Eq.~(\ref{eq:Nc_fraction}). This yields a nonlinear equation determining $p$ as
\begin{equation}
p = 2e^{-p \alpha t_{\rm r}} -1.
\end{equation}
Insertion of this result into Eq.~(\ref{eq:Nc_division}) yields the final result for the overall cell number as 
\begin{equation}
N(t) = N_{\mathrm{c}}(0) e^{\alpha p t}/p.
\label{eq:N}
\end{equation} 
Thus, the effective growth rate of the cell number in simulations is given by $\alpha p$. Formula~(\ref{eq:N}) has no free parameters and fits the simulation results very well, see the inset of Fig.~\ref{bacterianumber}(c). 

\subsection{Pilus-mediated interactions determine local colony order}
\label{ordering}
\begin{figure}[!htp]
	\includegraphics[width=0.95\columnwidth]{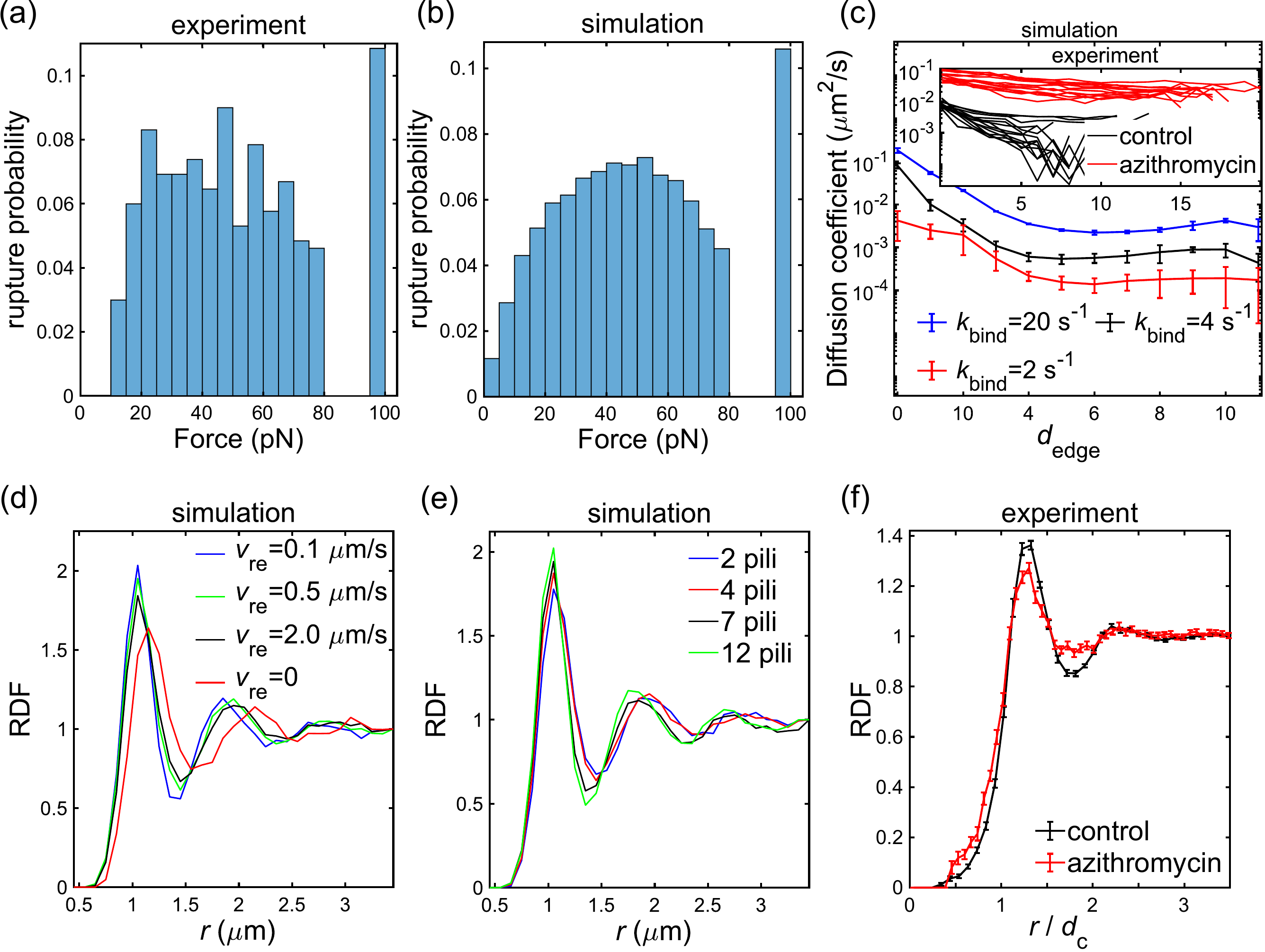}
	\caption{Pilus-generated forces and local colony order. (a)~Experimental results for the distribution of pilus-pilus bond rupture forces measured with an optical trap~\cite{zollner2019type}. Forces above 80pN cannot be measured precisely, the histogram bar at 100pN represents the contribution of all forces over 80pN.  (b)~Distribution of pilus-pilus bond rupture forces in simulations. 
	(c)~Diffusion coefficient of bacteria as a function of distance from the colony edge $d_{\rm edge}$ in simulations. Error bars show standard deviations in 3 samples. Inset: experimental data for wild-type cells and azithromycin-treated cells~\cite{cronenberg2021antibiotics}.
	(d)~Radial distribution function (RDF) of cells inside simulated colonies for different pilus retraction velocities. The shape of the functions, with decreasing, quasi-periodic maxima resembles the RDF of a liquid. For passive pili ($v_{\rm re}=0$), $k_{\rm bind}=2 \,{\rm s}^{-1}$ is used, compare Tab.~\ref{table02}. (e)~The maxima in the RDF become more pronounced with increasing numbers of pili per cell, thus, pili promote ordering. (f)~Experimentally determined RDF for wild-type cells and azithromycin-treated cells with fewer T4P~\cite{cronenberg2021antibiotics}. The cell diameter $d_{\rm c}$ is 1.02$\mu \mathrm{m}$ for the untreated control cells and is 1.42$\mu \mathrm{m}$ for the azithromycin-treated cells. Error bars are the standard error of the mean with a sample size of 24 colonies.
	}
	\label{RDF} 
\end{figure}

To establish that the parameter values chosen for simulating pilus dynamics and forces correspond to measured values for \textit{N.~gonorrhoeae}, the distributions of rupture forces in our simulations are recorded. For adjusting the parameters governing pilus binding and rupture, simulation results are compared with measured rupture force distributions, see Fig.~\ref{RDF}(a,b). The rupture-force values used for Fig.~\ref{RDF}(a) correspond to previously published data~\cite{welker2018molecular}, where the experimental procedures are explained in detail. To examine how the active force generation affects the mobility of cells in colonies, we next compare the simulated long-time diffusion coefficient of cells in colonies with stationary size with experimental data~\cite{cronenberg2021antibiotics}, see Fig.~\ref{RDF}(c). Consistent with the experimental results and previous computational work~\cite{ponisch2018pili, cronenberg2021antibiotics}, diffusive motion of cells decreases at the center of the colonies. These gradients in mobility are not a result of graded mechanical activity because all cells in the simulations have the same properties. Rather, the reduced motion inside the colonies is due to a ``caging'' of every cell by its neighbors~\cite{hennes2022caging}. This mutual obstruction of movement is reduced at the periphery of the colony as a result of the lower cell density. Treatment of the colonies with the antibiotic azithromycin reduces the T4P-T4P binding among neighboring cells. Accordingly, cell motility in colonies treated with azithromycin is increased, Fig.~\ref{RDF}(c,inset)~\cite{cronenberg2021antibiotics}. Note that the employed concentrations of azithromycin do not completely abolish T4P retraction or lead to a high cell death rate. In simulations, the reduced pilus interaction of azithromycin-treated cells are represented by variation of the binding constant for T4P, $k_{\rm bind}$, Fig.~\ref{RDF}(c).

Next, the local order in simulated colonies is investigated. The degree of local ordering is characterized by the radial distribution function (RDF), which is the average local particle density at distance $r$ from any reference particle, normalized by the average particle density of the system~\cite{younge2004model, kopera2018computing}. The RDF is defined as
\begin{equation}
g(r)=\frac{V}{N}\sum_{i=1}^{N}\frac{\phi_{i}(r)}{N V_{\rm shell}(r)},
\end{equation}
where $\phi_{i}(r)$ is the number of particles whose distance to the $i$th particle is between $r-\Delta r$ and $r+\Delta r$ with $\Delta r=0.05\,\mu\mathrm{m}$, $V_{\rm shell}(r)$ is the volume of the shell between radii $r-\Delta r$ and $r+\Delta r$, $N$ is the total number of particles in the system, and $V$ is the volume of the colony. In Fig.~\ref{RDF}(d,e), the RDFs of bacteria inside stationary, non-growing colonies are displayed. For cells carrying $2-12$ pili, which corresponds to the experimentally established number for wild-type \textit{N.~gonorrhoeae}, the RDFs have the typical characteristics seen for liquids with multiple, maxima that are decreasing in magnitude with increasing $r$. Hence, pilus-based cell-cell interaction generates structures with short-range order. Since the pili also cause relative motion of the bacteria, decreasing pilus retraction speed increases the spatial ordering as can be seen in Fig.~\ref{RDF}(d). Higher numbers of pili result in more pronounced maxima and therefore to a higher degree of spatial ordering, Fig.~\ref{RDF}(e). Experimentally, a lower number of T4P can be induced by treatment of the colonies with sub-inhibitory  concentrations of antibiotics~\cite{stephens1984loss, cronenberg2021antibiotics}. Figure~\ref{RDF}(f) displays experimentally measured RDFs for wild-type cells and azithromycin-treated cells. Distances in this plot are scaled by the different mean diameters of the bacteria. Lowering the number of T4P by azithromycin treatment reduces the local ordering, which is in good qualitative agreement with the simulation results. Experiments were performed as described previously~\cite{cronenberg2021antibiotics}.

\subsection{Active phase segregation in mixed colonies}
\label{sorting}

Experimentally, strains carrying mutations affecting the T4P machinery have been found to segregate during formation of colonies~\cite{oldewurtel2015differential,ponisch2018pili}. 
Bacterial segregation was seen to be dependent on the number of pili per cell, on post-translational pilus modifications that modify binding properties, and on the ability of bacteria to retract their pili. The observed colony morphotypes were suggested to be in agreement with the so-called differential-strength-of-adhesion hypothesis~\cite{harris1976cell}, which proposes that contractive activity of cells in addition to differential adhesiveness drives cell sorting. While active force generation was seen to be necessary for defined morphologies of mixed microcolonies, an experimental separation of the effect of pilus activity from differential adhesiveness is challenging due to the molecular complexity of pili. Simulations allow the systematic study of how variation of different parameters affects segregation. 

\begin{figure}[!htp]
	\includegraphics[width=\columnwidth]{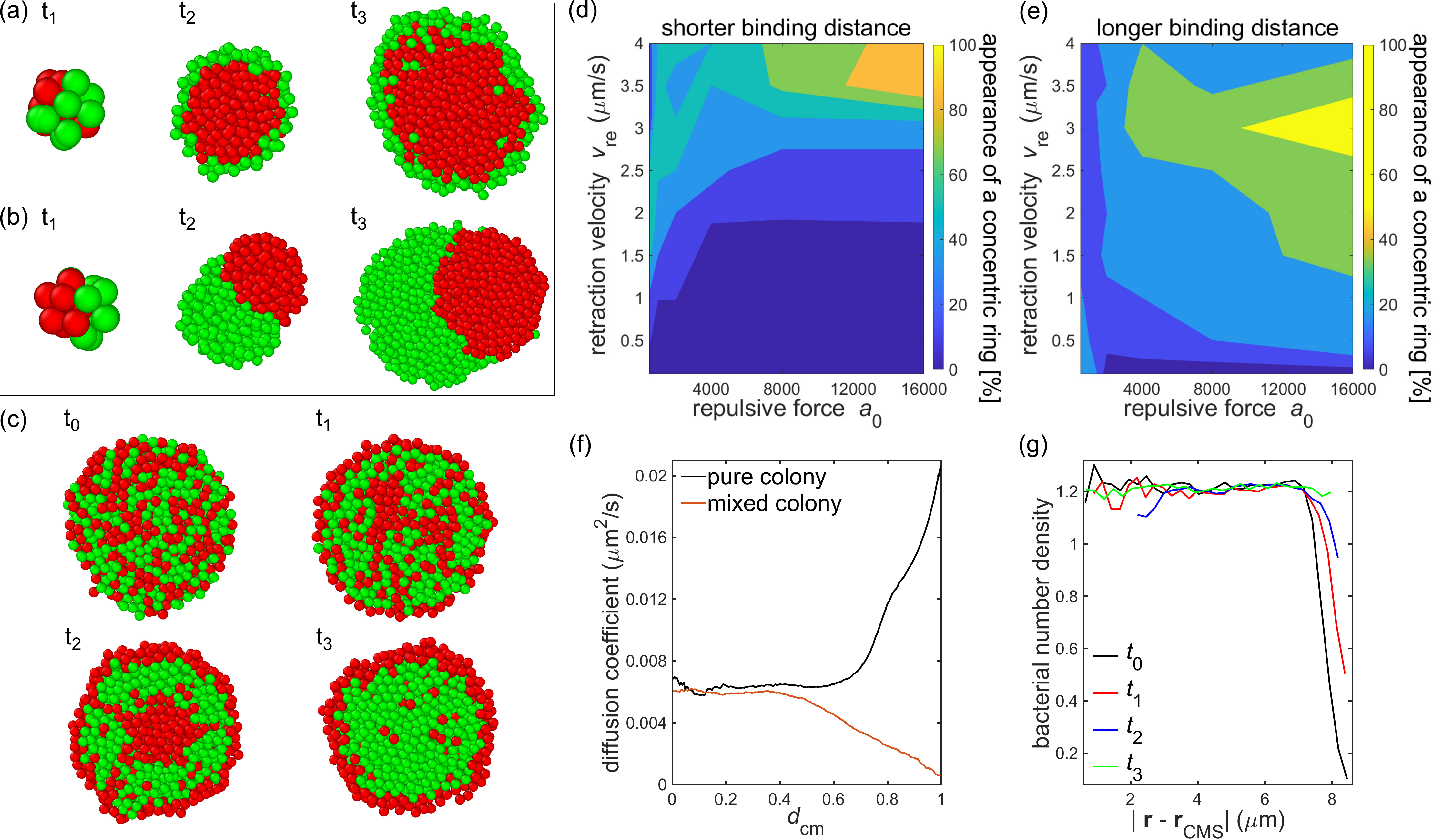}
	  \caption{Caption next page.}
		\label{mixture} 
\end{figure}

To establish that the simulations produce results that are consistent with experimental data, experimentally studied cases of colony segregation are re-investigated. We first simulate simultaneous growth of two kinds of strains carrying different numbers of pili, which is similar to earlier experimental work~\cite{oldewurtel2015differential}. In simulations, the growing colonies segregate and the cells that have many pili concentrate in the center of the colony, while cells with fewer pili form a spherical shell in the periphery, see Fig.~\ref{mixture}(a). Qualitatively, this configuration can be explained by the hierarchy of interaction strengths~\cite{oldewurtel2015differential}. The mutual attraction of a pair of cells with many pili is larger than the attraction of a cell with many pili to a cell with few pili. The weakest attraction occurs among pairs of cells with few pili. The formation a shell of weakly-binding cells in the periphery is energetically advantageous because of the reduction of surface-energy cost. For binary mixtures of bacteria with different pilus-rupture probabilities, other hierarchies of interaction strength are possible. For a mixture of two cell types that have high pilus rupture forces among each other, but lower rupture forces for pairs of different cells, simulations show the formation of two segregated half-spheres during growth, see Fig.~\ref{mixture}(b). This is consistent with experimental results, where wild-type cells were mixed with mutants deficient in post-translational pilin glycosylation~\cite{oldewurtel2015differential}.

Previous work on pilus-driven self-assembly of colonies has shown that binary cell mixtures consisting of cells with intact and retraction-deficient pili segregate~\cite{ponisch2017multiscale, ponisch2018pili}. To learn more about the segregation dynamics in this case, we start our simulations with fully grown colonies consisting of random binary cell mixtures, see Fig.~\ref{mixture}(c). Half of the cells can retract their pili and the other half are retraction-deficient. Since both cell types have the same number of pili, no differential in adhesiveness exists. Nevertheless, the initially random distribution of different cell types gradually disappears and the retraction-deficient cells accumulate at the periphery of the simulated colony, as found in earlier work~\cite{ponisch2017multiscale}. Unexpectedly, our simulations also predict the existence of a metastable intermediate state, in which active, pilus-retracting cells form a concentric spherical shell inside the colony, see Fig.~\ref{mixture}(c, $t_2$). This intermediate state has to our knowledge not yet been observed experimentally. The lifetime of the predicted intermediate state depends on the pilus-based interactions and on the strength of the cell-cell repulsion. We quantify the effect of pilus-retraction velocity and cell-cell repulsion on the appearance of the metastable concentric shell. Heat-maps of the average lifetimes of the concentric spheres as a percentage of the simulation time are shown in Fig.~\ref{mixture}(d). For short-ranged pilus interactions, the concentric shell of retracting cells inside the colony hardly appears, see Fig.~\ref{mixture}(d), ($d_{\rm bind}=2$). Likewise, this metastable state is suppressed if pilus-mediated interactions are limited to the next neighbors via Voronoi tesselation. The appearance of the metastable state requires long-ranged pilus-pilus interactions and rather stiff repulsive potentials among cells, see Fig.~\ref{mixture}(e) ($d_{\rm bind}=3$). Through such long-ranged pilus-pilus interactions, cells can exert forces on other cells that are not their direct neighbors. Experimentally, it has been established that the length of T4P follows an exponential distribution with a length scale around $0.8\,\mu \mathrm{m}$ and measured maximum lengths up to $5\,\mu \mathrm{m}$~\cite{kraus2022external}. Thus, the average T4P length is about the diameter of a coccus. For our simulations, we therefore chose a default binding cutoff equal to $1.5$ times the cell diameter plus two times the cell radius ($d_{\rm bind}=2.5$). Experimentally, longer-ranged interaction forces could occur for bacteria under stress conditions that affect T4P dynamics~\cite{kraus2022external} or in situations where extracellular matrix constituents, such as polysaccarides or DNA, transmit forces inside colonies. The simulation results suggest that an experimental observation of a metastable concentric shell during phase separation would point toward the existence of such long-ranged, pilus-based interactions among bacteria.

Figure~\ref{mixture}(f) shows plots of the diffusion coefficient of cells as a function of the distance from the colony center. For cell colonies consisting of one cell type, diffusive motion of cells decreases at the center of the colonies as already shown above. In contrast, for segregated colonies consisting of cells with retracting and non-retracting pili, the diffusion constant decreases with the distance from the colony center. This position-dependence of the cell mobility is consistent with the increasing concentration of pilus-retraction-deficient cells at the periphery of the colony, see Fig.~\ref{mixture}(g).

\subsection{Non-equilibrium fluctuations of colony boundaries}
\label{motion}
The position fluctuations of cells in colonies on the one hand provide information about the viscoelastic properties of the system and, on the other hand, carry information about the non-equilibrium forces holding the system together. While it is trivial to keep track of cell positions in simulations, a high-precision measurement of cell positions inside a three-dimensional colony is challenging in experiments. However, it is possible to image whole colonies with high frame rate and subsequently extract the colony edges from the images. Thus, the non-equilibrium fluctuations of colony boundaries are observable. We mimic here such a measurement in simulations by tracking cells located in a fixed small sector at the edge of colonies in the stationary state, as shown in Fig.~\ref{psd}(a). In this setup, a movement of bateria at the colony edge can either result from thermal noise or the activities of pili. Colonies are grown in simulations using wild-type cells. After switching off colony growth, the role of pilus activity for the stationary state is studied. The decay time of the velocity autocorrelation function (VACF) in the stationary state without active pilus-mediated forces, $v_{\mathrm{re}}=0$, corresponds to the inertial time scale in simulations. Thus, we roughly have an inertial decay time $t_{\mathrm{inert}} \simeq 0.1\,k^{-1}_{\mathrm{rupt}}$ (simulation units), see Fig.~\ref{psd}(b). The cell motion resulting from pilus retraction strongly increases the VACF below the time scale of pilus-bond rupture $k^{-1}_{\mathrm{rupt}} = 1\,\mathrm{s}$. 

With the radial distance $r_{\mathrm{CMS}}(t)$ between the center of mass and the edge of the colony, the deviations from the time average are given by $X(t)=(r_{\mathrm{CMS}}(t)-\langle r_{\mathrm{CMS}} \rangle)$. The power spectral density (PSD) of the displacement is given by 
\begin{equation}
P(\omega) = \frac { |\widetilde{X}(\omega)|^2}{s\, n},
\label{psdeq}
\end{equation}
where $\widetilde{X}(\omega)$ is the discrete Fourier transform of $X(t)$, $s$ is the sampling rate, and $n$ is the number of data points. The PSDs of the radial motion of bacteria at the boundary in our simulated colonies are shown as Fig.~\ref{psd}(c). Fluctuations with frequencies $\omega \lesssim 10 \,\mathrm{Hz}$ are expected to be experimentally accessible. For $\omega>2 \pi/t_{\mathrm{inert}} \simeq 50\,\mathrm{Hz}$, the results are not expected to match with experiments since here inertial effects start to play a role the in simulations.

First, retraction-deficient, passive pili with $v_{\mathrm{re}} = 0$ are considered. Since these pili only form temporary, rupturing bonds between the cells, they produce an effective friction among cells. The boundary fluctuation are similar to the motion of an overdamped particle in a purely viscous environment $P(\omega) \propto \omega ^{-2}$. Second, for retraction-deficient, passive pili that form permanent bonds ($v_{\mathrm{re}} = 0$, no rupture), we find boundary fluctuations that are similar to the motion of an overdamped particle in an harmonic potential with $P(\omega) \propto \mathrm{const.}$ at low frequencies and $P(\omega) \propto \omega ^{-2}$ for high frequency (not shown). Third, wild-type cells with retracting T4P are considered ($v_{\mathrm{re}} = [0.5,2]\,\mu\mathrm{m}$/s). In this case, pilus retraction on the one hand enhances the elastic forces among cells, on the other hand, increases the rupture rate of bonds formed by T4P. Overall, the activity of T4P results in visco-elastic material properties with a pronounced elastic response at low frequencies. Simulations show that this elastic response does not occur if temporary cell-cell connections are formed by passive links.

The inset of Fig.~\ref{psd}(c) show experimental results for the PSDs
of wild-type cells and a strain carrying inactivating deletions in genes that encode the phosphotransferase \textit{pptA}, which required for the post-translational modification of T4P~($\mathit{\Delta}$\textit{pptA}). For wild-type cells, where colony contains around 4000 cells, the experimentally measured colony boundary fluctuations up to $10$~Hz are in good qualitative agreement with our simulation results. For a qualitative assessment of the experimental data for the $\mathit{\Delta}$\textit{pptA} strain, a smaller colony with about 600 mutants is simulated. Here, the parameters governing T4P dynamics are adjusted to mimic the higher binding probablility and lower rupture frequency measured for the $\mathit{\Delta}$\textit{pptA} strain in comparison to the wild-type~\cite{zollner2019type}. A binding rate of $k_{\rm bind}^{\rm mutant}=50\,{\rm s}^{-1}$ and lower rupture rate $k_{\rm rupt}^{\rm mutant}=1\,{\rm s}^{-1}$ were chosen, compare Tab.~\ref{table02}. We also find that a lower retraction velocity needs to be chosen for the simulated mutant strain, compared to the wild-type strain ($v_{\mathrm{re}} = 0.5\,\mu\mathrm{m}/\mathrm{s}$). This lower retraction velocity is necessary in simulations to mimic the lower retraction frequency of the mutant~\cite{zollner2019type} and reduces the apparent elastic modulus at low frequencies. Overall, the higher plateau value of the PSD at low frequencies suggest that post-translational modification of T4Ps in wild-type cells leads to ``stiffer'' \textit{N.~gonorrhoeae} colonies. 

Since shape fluctuations of a wild-type cell colony mainly result from active forces, a violation of the equilibrium fluctuation-response relation is expected. To find out how a fluctuation-response relation can be measured experimentally, we simulate a setup for controlled mechanical perturbation of the colony boundary. This setup is inspired by techniques for measuring active fluctuations in cell membranes~\cite{turlier2016equilibrium}. We fix a simulated colony between walls and stick a bead with radius $R_{\rm B}=1.5 \,\mu {\rm m}$ onto one side of the colony, see Fig.~\ref{psd}(d). The same parameter values are used to describe pilus interaction with walls and pilus-pilus interaction. 
The pairwise interactions between the bead and the cells is modeled with a Morse potential. Denoting the distance between the bead and any neighboring cell $i$ by $r_{i,\mathrm{B}}$, the potential is given by
\begin{align}
	\Psi^{B}_i(r_{i,\mathrm{B}}) &= c_{\mathrm{mors}}\,[e^{-2 \beta (r_{i,\mathrm{B}}-R_{i,\mathrm{B}})} - 2 e^{- \beta (r_{i,\mathrm{B}}-R_{i,\mathrm{B}})}],& & \mathrm{for}\;\; r_{i,\mathrm{B}} \leq d_{\mathrm{mors}},\\
	\Psi^{B}_i(r_{i,\mathrm{B}}) &= \Psi^{B}_i(d_{\mathrm{mors}}),& & \mathrm{for}\;\; r_{i,\mathrm{B}} > d_{\mathrm{mors}},
\end{align}
where sum of the radii of cell and beads is given by $R_{i,\mathrm{B}}= R+R_{\rm B}$. The cutoff for the interaction potential is set at $d_{\mathrm{mors}} = 3.1 \,\mu {\rm m}$. Other parameter values of the potential are fixed as $c_{\mathrm{mors}}=10 \,\,\mathrm{pN} \mu {\rm m}$ (energy unit: $f_c d_c$) and $\beta = 1 \,\mu {\rm m} ^{-1} $. The radial displacement of the bead relative to the center of the colony, $x(t)$, is employed to quantify the fluctuations of the colony boundary through a PSD $P(\omega)$ given by Eq.~(\ref{psdeq}). Alternatively, a sinusoidally varying force $F_{\mathrm{ext}}(t)$ is applied to the beads' center, pointing towards the colony center. The Fourier transform of the force is given by $\hat{F}_{\mathrm{ext}}(\omega)$ with angular frequency $\omega$. The response function is given by
\begin{eqnarray}
	\hat{\chi}(\omega) &\equiv& \frac{\hat{x}(\omega)}{\hat{F}_{\mathrm{ext}}(\omega)}.
\end{eqnarray}
The imaginary part of the response function $\hat{\chi}(\omega)$ is denoted by $\hat{\chi}'(\omega)$ and we define the quantity $H(\omega) \equiv -\hat{\chi}'(\omega)2k_B T/\omega$. For systems in thermal equilibrium, the fluctuation-response theorem states that
\begin{eqnarray}
P(\omega) = H(\omega).
\label{eq:fdr}
\end{eqnarray}
For colonies consisting of bacteria with retraction-deficient, passive pili and permanent bounds ($v_{\mathrm{re}} = 0$), Eq.~(\ref{eq:fdr}) is satisfied and the fluctuation-response theorem holds as expected, see Fig.~\ref{psd}(e). In simulations of colonies consisting of wild-type bacteria that can retract their pili ($v_{\mathrm{re}} = 1.0\,\mu\mathrm{m}/\mathrm{s}$) and form dynamic bonds with other cells, the equilibrium fluctuation response relationship is violated across the whole experimentally relevant frequency range of $[0-10]\,$Hz. Note that constraining the colony in between walls and then tracking the motion of a bead is not equivalent to tracking the distance of the colony boundary from its center. Hence the spectral densities in Figs.~\ref{psd}(c),(f) are different. For wild-type bacteria, the simulations predict a very strong deviation from Eq.~(\ref{eq:fdr}), where $P(\omega)$ is several orders of magnitude larger than $H(\omega)$. Such deviations are likely measurable in experiments. 

\subsection{Active colony spreading on a surface}
\label{sec:spreading}
An important aspect of growing bacterial colonies is the colonization of surfaces and the invasion of tubes and channels. To investigate the role of active adhesion forces for the colony behavior in such situations, we consider the interaction of cells with walls that provide attachment sites for pili. Walls are represented by a layer of immobile, soft spheres. Cell-wall interactions are represented by the same conservative potential employed for cell-cell repulsion. Since biomolecular binding affinities are typically determined by the unbinding rate, the pilus unbinding rate $k_{\rm plane}$, corresponding to $k_{\rm rupt}$ for pilus-pilus unbinding, is varied. The other parameters governing  pilus-wall binding are assumed to be the same as for pilus-pilus interactions, see Tab.~\ref{table02}. Experimentally, such wall properties can be realized, e.g., by coating hydrogel surfaces with pilin.

For colonies spreading on a planar wall, the shape results from a competition between
the cell-cell interactions within the colony and the interactions of the cells with the substrate. Previous simulation studies showed that the radius of the contact zone between the colony and the wall increases with the rupture force scale~\cite{ponisch2017multiscale}, which can be called ``partial wetting''. Here, we vary the dissociation-rate constant of the pilus-wall bonds to assess the wetting transition. Simulation snapshots of colonies growing on a planar surface are shown in Fig.~\ref{surface}(a-c). If the dissociation-rate constant of the pilus-wall bonds is smaller than the dissociation rate constant for pilus-pilus bonds, $k_{\rm plane} \lesssim 2\,\mathrm{s}^{-1} < k_{\rm rupt} = 3\,\mathrm{s}^{-1}$, we find that the colonies dissolve and the bacteria are evenly dispersed along the surface, see Fig.~\ref{surface}(a,d), which corresponds to complete wetting. For $k_{\rm plane} \geq k_{\rm rupt}$, the colonies assume rounded shapes that can still remain in loose contact with the surface, see Fig.~\ref{surface}(b,c).\\
To assess the dynamics of the wetting process, we next record the diameter $d_{\mathrm{surface}}$ of the contact zone of a spreading colony on the surface. For a passive, Newtonian fluid on a planar surface, the diameter of a spreading droplet asymptotically obeys a power-law dependence on time $t$ as $d_{\mathrm{surface}}\sim t^{\vartheta}$, known as Tanner's law~\cite{voinov1976hydrodynamics,tanner1979spreading,bonn2009wetting}. The exponent $\vartheta$ depends on droplet size and on the dimension. Droplets that are much smaller than the capillary length obey in three dimensions for long times the scaling $\sim t^{1/10}$~\cite{bonn2009wetting}, which results from a leading-order balance of capillary forces with dissipation close to the wetting line. It has also been theoretically predicted that thermal fluctuations promote spreading of nanodroplets and lead to a scaling of $\sim t^{1/6}$~\cite{davidovitch2005spreading}. In our simulations of active bacterial colonies, a regime with the classical passive-liquid scaling $\sim t^{1/10}$ is not observed. Rather, we find that the diameter $d_{\mathrm{surface}}$ obeys a power law with an exponent close to $1/4$, which is very similar for different parameter choices, see Fig.~\ref{surface}(e). Such a scaling indicates that the dynamics is dominated by a balance of surface-attraction and dissipation in the bulk of the colony. The scaling breaks down at long times when the colony reaches a stationary, rounded shape on the surface. 

\subsection{Active colony invasion of narrow channels}
\label{sec:invasion}
We next simulate the invasion of small channels by colonies. The colonization of protective niches can present a selective advantage in abiotic environments and can also be an important aspect of host-pathogen interaction. Previous work on \textit{Neisseria meningitidis}, the causative agent of meningitis, showed that attractive forces generated by T4P fluidize the bacterial colonies, which is required for efficient colonization of the blood capillary network during infection. Furthermore, simulations of \textit{N.~gonorrhoeae} migration through asymmetric corrugated channels show a rectification of motion for active bacteria~\cite{bisht2020rectification}. However, a systematic assessment of the conditions necessary for the active invasion of constrictions is missing. To focus on the role of pilus activity, we only consider colonies that are not growing or dividing and channels are represented with the same methods as walls in the previous Section ``Active colony spreading on a surface''. 

Like for cell-surface interaction, the behavior of active colonies is seen to be qualitatively similar to a liquid minimizing surface energy. Active pilus retraction can cause a rapid and complete invasion of the channel, see Fig.~\ref{channel}(a-e). For passive cells ($v_{\mathrm{re}} = 0$), colonies can attach to the walls but proper invasion of the whole channel is not observed, see Fig.~\ref{channel}(f). A complete entrance of passive colonies into the channels never occurs in our simulations, even for large surface affinity, $k_{\rm plane} \simeq 0.01 \,\mathrm{s}^{-1}$, and very long simulation times. For active colonies, the onset of channel invasion occurs rather suddenly when increasing the affinity for the substrate ($\sim 1/k_{\mathrm{plane}}$), see Fig.~\ref{channel}(d). However, the threshold value of $k_{\mathrm{plane}}$ below which channel invasion occurs is not the same as the threshold required for complete wetting of a planar substrate shown in Fig.~\ref{surface}. The reason for different threshold affinities is presumably that the formation of a monolayer of cells on a planar surface is energetically more costly than formation of a cylindrical colony with finite internal volume. Consistent with this interpretation, we find that the narrower the channel is, the higher the surface-binding affinity has to be to achieve channel invasion, see Fig.~\ref{channel}(g). For very narrow channels, $w \lesssim 2.5 \, \mu \mathrm{m}$, we find that the invasion does not occur through collective motion of an intact colony but that individual cells and small collections of cells break off from the colony and individually explore the channel, see Fig.~\ref{channel}(e). A possible cause for this break-up is that the high curvature of very narrow channels results in a surface area per cell that is larger than the surface area per cell on a plane. Since the number of T4P is limited, the geometry of narrow channels increases the effective binding affinity between cells and walls and decreases the effective binding affinity among cells is reduced. The break-up of colonies during invasion of these channels is therefore due to the finite inherent length scale of the ``bacterial active fluid''.\\
To quantify the dynamics of colony invasion, we next record the speed of the front of the colony moving down the channel and plot it as a function of the enter length $L(t)$, see Fig.~\ref{channel}(h). For passive liquids, the penetration dynamics into horizontal capillary tubes under the assumption of negligible gravity and inertia obeys an approximate scaling of $L(t) \sim \sqrt{t}$, which is derived as follows~\cite{martic2002molecular}. The liquid viscosity is denoted by $\eta_{\mathrm{p}}$, the presumably constant surface-contact angle is $\theta_{\mathrm{p}}$, the surface tension is $\sigma_{\mathrm{p}}$ and the channel diameter $w$. Then, the balance of capillary driving force with viscous friction can be written as $8 \eta_{\mathrm{p}} L(t)\dot{L}(t)/w^2 = \sigma_{\mathrm{p}} \cos{\theta_{\mathrm{p}}}/w$. Solution of this differential equation for $L(t)$ yields the Lucas-Washburn equation~\cite{lucas1918ueber,washburn1921dynamics}
\begin{align}
    L(t) = \sqrt{\frac{w\,\sigma_{\mathrm{p}} \cos{\theta_{\mathrm{p}}}}{8\,\eta_{\mathrm{p}}}}\sqrt{t}.
\end{align}
For our active colonies, we find that the invasion dynamics for thin channels of width $w = 3.3 \, \mu \mathrm{m}$ obey the $L(t) \sim \sqrt{t}$ scaling of passive liquids. These channels are wide enough to prevent colony break-up. For thicker channels the Lucas-Washburn-like scaling no longer holds in our simulations, as is the case for passive liquids, where the deviations are attributed to inertial effects and to a dependence of the contact angle on the wetting dynamics.

Overall, channel invasion by active bacterial colonies displays a striking qualitative similarity to capillary wetting by passive liquids. However, for passive colonies, having a mesoscopic internal length scale, we do not observe channel invasion in simulations. The main role of pilus-mediated random activity is to increase the fluidity and to thereby change the dynamics of colony-surface interactions. Thus, T4P activity allows the occurrence of channel invasion and surface spreading on biologically relevant time scales.

\section{Conclusions}
\label{sec:conclusion}
We have introduced a simulation model to study the non-equilibrium structure and dynamics of colonies of active, growing bacteria on different time- and length scales. Bacterial cells are modeled with an algorithm akin to dissipative particle dynamics. Parameter values are carefully chosen to allow comparison of the simulation results with experimental measurements. We investigate different physical aspects of \textit{N.~gonorrhoeae} colonies, including growth dynamics, local ordering, and self-sorting of bacteria in colonies. Simulation results are in good qualitative agreement with experimental data. We also propose a setup for measuring fluctuations in the colony shape and its response to external force. The simulations predict a strong, measurable violation of the equilibrium fluctuation-response relationship. Furthermore, the model shows that actively fluctuating adhesion forces can allow the bacterial invasion of narrow channels. Thus, active force generation is not only required for bacterial migration, but can determine the rheology of cell colonies and drive the colonization of constricted environments, which represent central aspects for host infection and bacterial contamination of abiotic environments. At present, basic physical mechanisms underlying the collective interaction of active particles with complex surfaces are hardly understood. We expect that future experimental and theoretical work on the non-equilibrium properties of bacterial colonies will generate insights that deepen our understanding of the emergent properties of such active matter systems.

\section{Methods}
\subsection{Simulation details and parameter values}
The simulation code is integrated into the molecular dynamics simulator LAMMPS~\cite{plimpton1995fast}, which allows an efficient parallelization while providing great flexibility regarding the model choice. To model the cellular dynamics described above, we wrote a C++ code. The velocity-Verlet algorithm is used to advance the set of positions, velocities and forces. The code is parallelized for execution on CPUs and large colonies consisting of ten-thousands of bacteria can be simulated efficiently. The colonies are visualized with OVITO~\cite{stukowski2009visualization}.

The characteristic scales that are used as simulation units are the cell diameter $d_{\rm c} = 2 R = 1\,\mu \mathrm{m}$, a time scale given by the inverse of the default value of the pilus-unbinding rate constant $t_{\rm c}=1/k_{\mathrm{rupt}}=1 \mathrm{s}$, and a force scale of $f_{\rm c} = 1\,\mathrm{pN}$. Parameters values that are used for the simulations are listed in Tab.~\ref{table02}. Whenever alternative parameter values are used, they are provided with the results.

\begin{table}
	\centering
	\caption{The choice of parameters for the simulations.}
	\begin{tabular}{llll}
		\hline 
		Parameter & Value & Unit & Reference \\
		\hline 
		cell radius $R$ &  0.5 & $d_{\rm c}$ & \cite{welker2018molecular}\\
		cell mass $m$ & 0.1 & $f_{\rm c} t_{\rm c}^2 d_{\rm c}^{-1}$ \\
		pilus spring constant $k$ & 500  & $f_{\rm c} d_{\rm c}^{-1}$ & \\ 
		pilus stall force $f_s$ & 180 & $f_{\rm c}$ & \\
		maximum retraction velocity of pili $v_0$  & 2 &$d_{\rm c} t_{\rm c}^{-1}$ & \\
		number of pili per cell & 7 & &  \cite{holz2010multiple} \\ 
		simulation time step $\Delta t$ & $1\times10^{-4} $ &  $t_{\rm c}$ &\\
		division rate $\alpha$ & 1/500 & $t_{\rm c}^{-1}$ &\\ 
		diplococcus growth parameter $\nu_{\mathrm{r}}$ & 1.0  & $d_{\rm c}^{-1}$\\
		pilus characteristic rupture time $t_1$ & 0.5 & $t_{\rm c}$\\
		pilus characteristic rupture time $t_2$ & 0.091 & $t_{\rm c}$\\
		pilus characteristic rupture force $F_{\rm {c,1}}$ & 4.5 & $f_{\rm c}$ & \\
		pilus characteristic rupture force $F_{\rm {c,2}}$ & 60 & $f_{\rm c}$ & \\
		pilus rupture rate $k_{\rm {rupt}}$ & 3 & $t_{\rm c}^{-1}$\\
		pilus binding cutoff distance $d_{\rm {bind}}$ & 2.5 & $d_{\rm c}$\\
		pilus binding rate $k_{\rm {bind}}$ & 10 & $t_{\rm c}^{-1}$\\
		pilus-pilus bond rupture force scale $F_{\rm {rupt}}$ & 22.5 & $f_{\rm c}$ & \cite{welker2018molecular,oldewurtel2015differential}\\
		maximum conservative force $a_{0}$ & 4000 & $f_{\rm c} d_{\rm c}^{-1}$ \\ 
		conservative force cutoff $d_{\mathrm{con}}=2R$ & 1.0 & $d_{\rm c}$\\
		diplococcus spring constant $a_{\rm {growth}}$ & 4000 & $f_{\rm c} d_{\rm c}^{-1}$ &\\
		friction coefficient $\gamma$ & 50 & $f_{\rm c} t_{\rm c} d_{\rm c}^{-1}$ \\
		thermal energy scale $ k_{\rm B}T$ & $1\times10^{-5}$  & $f_{\rm c} d_{\rm c}$ \\
		dissipative and random force cutoff $d_{\rm {dpd}}$ & 1.5 & $d_{\rm c}$\\ 
		
		\hline 
	\end{tabular}
	\label{table02}
\end{table}

\subsection{Separation of time scales}
In the simulations, the time scale of viscous relaxation is smaller than the time-scale of pilus-based interaction. Thus, inertial effects are negligible. Moreover, the time scale of the pilus-based interaction is much smaller than the time-scale of cell division $t_{\rm c} \ll 1/\alpha$. Simulations typically start with one bacterium and colonies are formed by letting the cells grow and divide. The colony structures emerge during growth due to the repulsive interactions, thermal noise, and pilus-based interactions, as shown in Fig.~\ref{bacterianumber}(a). In some simulations, it is desirable to completely remove the effect of cell growth on the bacterial dynamics. For this purpose, cell growth and division are switched off after a sufficient colony size is reached. 

\subsection{Experiments}
The presented data, with the exception of the data for Fig.~5c, was originally generated for earlier experimental work~\cite{zollner2019type,cronenberg2021antibiotics,hennes2022caging}. Bacterial colonies were grown as described previously~\cite{stephens1984loss,zollner2019type}. Briefly, for assessing colony structure and dynamics, bacteria were incubated within a flow chamber under continuous nutrient supply for one hour to several hours hours. Constant supply of nutrients and, if used, antibiotics was ensured by applying continuous flow. For calculation of the RDF, bacteria were stained with Syto 9 to enable detection of the position of individual cells and to determine the cell volume. Colony dynamics were assessed with gfp and mcherry expressing cells~\cite{zollner2019type}. The displayed experimental data relates in detail to previous work as follows. The confocal section of a microcolony of fluorescently labeled bacteria in Fig.~\ref{bacterianumber}b) was produced as described in an experimental study on the effect of caging cellular motion in colonies~\cite{hennes2022caging}. The rupture forces of T4P bonds shown in Fig.~\ref{RDF}a) were measured with an optical trap~\cite{zollner2019type}. The diffusion coefficients and RDFs shown in Figs.~\ref{RDF}c),f) were measured as described for an experimental study of the effect of antibiotics on colony morphology~\cite{cronenberg2021antibiotics}. The PSDs of the colony boundary fluctuations shown in Fig.~\ref{psd}(c) were calculated from time-lapse images  of colonies recorded at 10~Hz for one minute. The experimental methods for recording the time-lapse images are detailed in earlier reports of experimental work~\cite{zollner2019type, welker2018molecular}.

\section*{Data availability}
Raw data for the presented figures will be provided upon reasonable request to the authors.

\section*{Code availability}
The code developed in this work is available upon reasonable request to the authors.

\section*{Competing interests}
The authors declare that there are no competing interests.

\section*{Author contributions}
KZ, GG, and BS performed theoretical work. MH and BM performed experimental research. All authors contributed to the writing of the manuscript.

\begin{acknowledgments}
KZ acknowledges kind support from the China Scholarship Council (CSC, No. 201804910439). BM acknowledges support by the Deutsche Forschungsgemeinschaft through grant MA3898. Funding by the European Research Council through a starting grant for BS is acknowledged (BacForce, g.a.No.~852585).
\end{acknowledgments}

%

\addtocounter{figure}{-1}
\begin{figure} [t!]
	\caption{Phase segregation of mixed colonies. (a)~Simulation snapshots for three time points, $t_1 - t_3$, of a growing colony consisting of cells with 14 pili (red) and cells with 7 pili (green).
	(b)~Snapshots for three time points, $t_1 - t_3$, of the simulated growth of a mixture of wild-type cells (WT, green) and mutants deficient in post-translational pilin glycosylation (GD, red)~\cite{oldewurtel2015differential}. Pilus rupture forces are as follows: $F^{\mathrm{WT-WT}}_{\mathrm{rupt}}=22.5\,{\mathrm{pN}}$, $F^{\mathrm{GD-WT}}_{\mathrm{rupt}}=20\,{\mathrm{pN}}$, $F^{\mathrm{GD-GD}}_{\mathrm{rupt}}=45\,{\mathrm{pN}}$. In (a) and (b), Voronoi tessellation is used to locate neighbors for pilus binding. 
	(c)~Segregation of a binary mixture of wild-type bacteria (green) and pilus-retraction-deficient cells (red). Pilus-mediated interactions are long-ranged in this example with $d_{\rm bind}=2.5\,\mu{\mathrm{m}}$. At time $t_0$, the colony is randomly mixed. Over time, the proportion of pilus-retraction-deficient cells increases in the colony periphery, $t_1$, and wild-type cells then accumulate in a concentric sphere inside the colony, $t_2$. The concentric sphere eventually disappears and wild-type cells accumulate in the colony center, $t_3$. (d)~Lifetime of the concentric sphere arrangement for short-ranged pilus interactions, $d_{\mathrm{bind}}=2.5\,\mu{\mathrm{m}}$. Lifetimes are given in percent of the longest observed lifetime ($3000$s). (e)~Lifetime of the concentric sphere arrangement for long-ranged pilus interactions, $d_{\rm bind}=3.5\,\mu{\mathrm{m}}$. Lifetimes are given in percent of $3000\,\mathrm{s}$. (f)~Diffusion coefficients of individual cells as a function of their distance from the colony center divided by the colony radius, $d_{\mathrm{cm}}$. Pure colonies consist of one type of cells, the mixed colony is the segregated system shown in (c, $t_3$). (g)~The mean radial number density of wild-type cells (green) for the simulation snapshots shown in (c).}
	\label{mixture-caption} 
\end{figure}

\begin{figure}[htp]
	\includegraphics[width=1.0\linewidth]{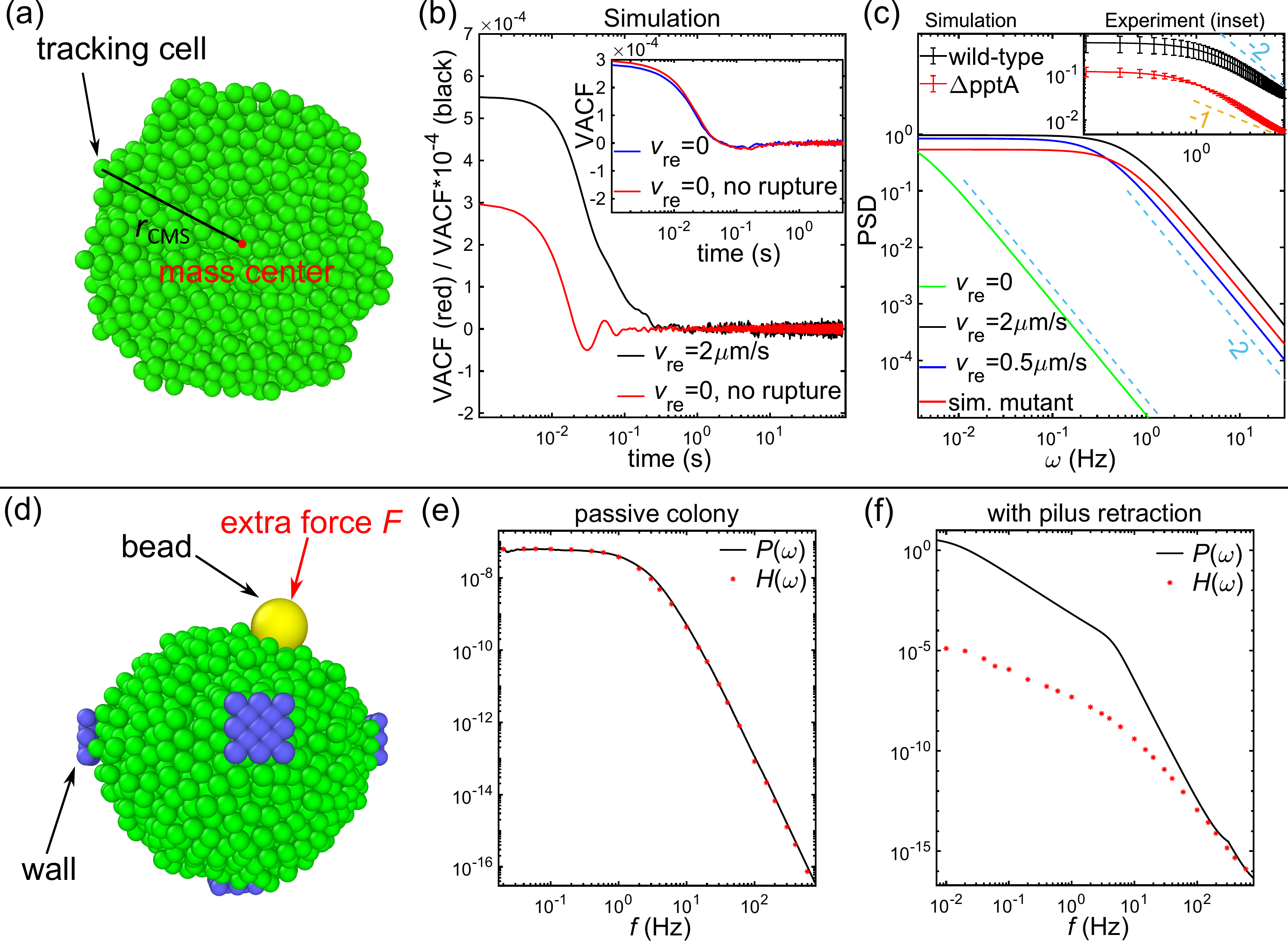}
	\caption{Caption next page.}
	\label{psd} 
\end{figure}		
\addtocounter{figure}{-1}
\begin{figure} [t!]
	\caption{
	Active fluctuations at the  periphery of colonies. (a)~Simulated setup for quantifying boundary fluctuations by measuring the radial distance $r_{\mathrm{CMS}}$ between a fixed angular position on the surface and the colony center.
	(b)~Active pilus retraction results in a slower decay of the velocity autocorrelation function (VACF) of the surface point. Main plot: colonies are first grown from cells with retracting pili and the role of pili is studied after growth is switched off. Inset: colonies are grown with retraction-deficient cells. (c)~Active pilus retraction produces a power spectral density of fluctuations characteristic for a visco-elastic material with an elastic behavior at low frequencies. For passive colonies, $v_{\mathrm{re}} = 0$, bond rupture results in a viscous material behavior. Colony size is 4000 cells. For comparison with experimental data, a small colony with 600 mutant cells is simulated having a higher binding rate $k_{\rm bind}^{\rm mutant}=50\,{\rm s}^{-1}$, lower rupture rate $k_{\rm rupt}^{\rm mutant}=1\,{\rm s}^{-1}$ and lower retraction velocity $v_{\mathrm{re}} = 0.5\,\mu\mathrm{m}/\mathrm{s}$. 
	Inset: experimental data for wild-type cells and a $\mathit{\Delta}$\textit{pptA} strain, error bars show standard deviations with 3 samples. (d)~Simulated setup for colony-shape perturbation. (e)~Simulated mutant colonies with retraction-deficient pili that form permanent bonds. The equilibrium fluctuation-response relationship holds. (f)~For simulated wild-type cells, the equilibrium fluctuation-response relationship is strongly violated ($v_{\mathrm{re}} = 1.0\,\mu\mathrm{m}/\mathrm{s}$, $d_{\mathrm{bind}}=3.0\,\mu{\mathrm{m}}$)}
	\label{psd-caption} 
\end{figure}

\begin{figure}[!b]
	\includegraphics[width=1.0\columnwidth]{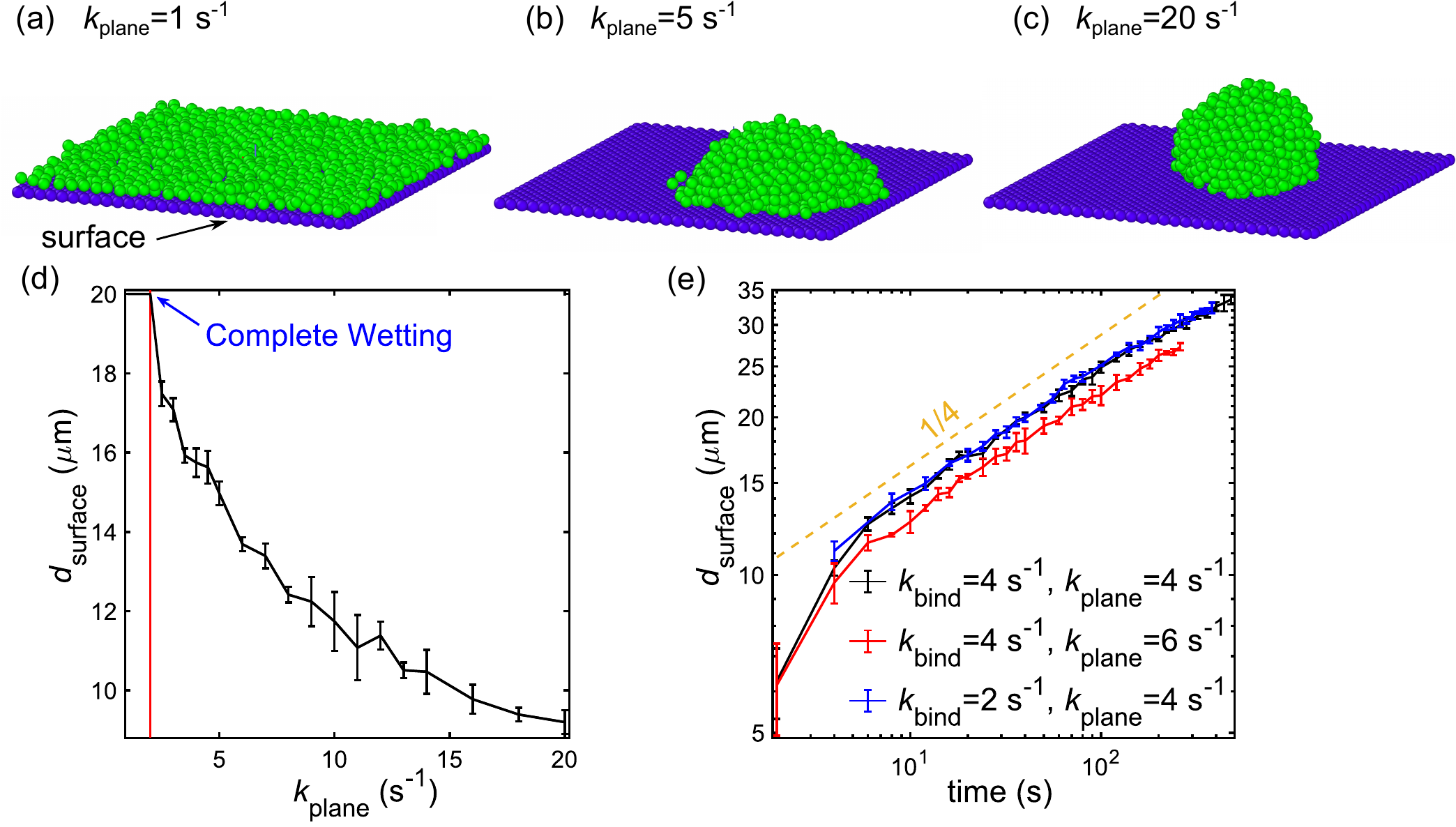}
	\caption{Colony spreading on planar surfaces. (a-c)~Simulation snapshots of spreading colonies. Depending on the dissociation rate constant of pilus-surface bonds, $k_{\rm plane}$, colonies undergo a partial or complete wetting transition. (d)~A complete wetting occurs when $k_{\rm plane} \ll k_{\rm rupt} = 2\,\mathrm{s}^{-1}$. (e)~The time dependence of the diameter of the spreading colony, $d_{\mathrm{surface}}$, obeys approximately a power law. If not given otherwise, pilus-substrate binding rates and rupture forces are assumed to be the same as for pilus-pilus bonds, see Tab.~\ref{table02}. Error bars represent standard deviations from samples of 3 simulations.}
	\label{surface} 
\end{figure}

\begin{figure}[!t]
	\includegraphics[width=1.0\columnwidth]{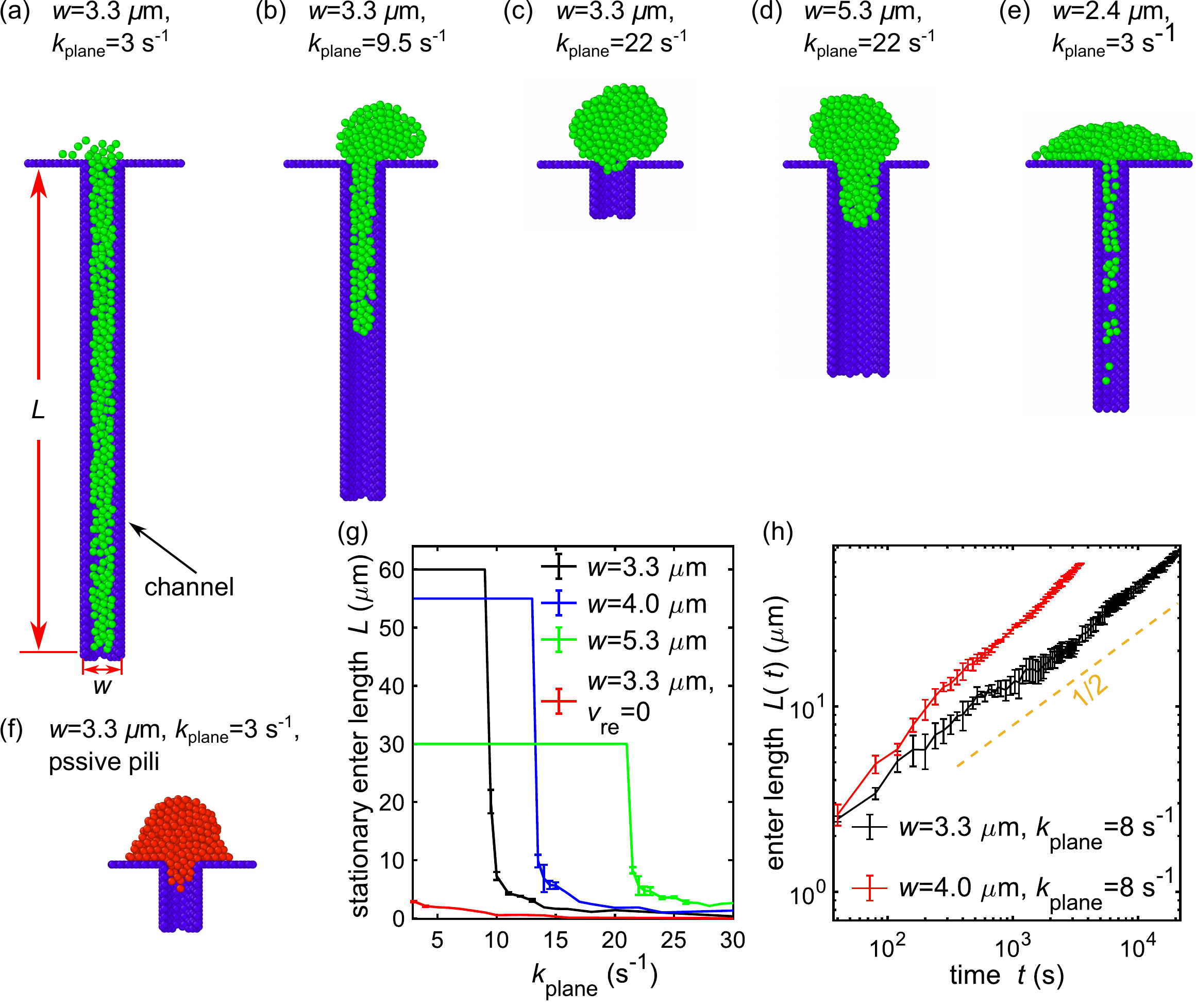}
	\caption{Colonies atop surfaces containing channels.  
	(a-c)~Simulated wild-type colonies that are initially positioned on top of a narrow channel can invade the channel by making use of pilus activity if the pilus-substrate bonds are strong. The length that a colony enters a channel is denoted by $L$. (d)~Colonies more easily invade a wider channel. (e)~Invasion of very narrow channels is possible, but colonies break up in this geometry (f)~Passive colonies are not seen to invade channels fully in simulations. (g)~Stationary lengths of colonies after entering channels of different widths. Depending on the channel width, full entrance of the colony occurs below a critical value of $k_{\mathrm{plane}}$. (h)~The time evolution of the enter length is described approximately by a power law. Error bars represent standard deviations from 3 simulation samples.
	}
	\label{channel} 
\end{figure}

\end{document}